\documentclass[preprint2]{aastex}

\begin{document}

\title{A Region of Violent Star Formation in the~Irr~Galaxy
IC~10: Structure and Kinematics of Ionized and Neutral Gas}

\author{O.V. Egorov \altaffilmark{1}, T.A. Lozinskaya \altaffilmark{1},
 A.V. Moiseev \altaffilmark{2}}

\altaffiltext{1}{Sternberg Astronomical Institute,
Universitetskiy pr. 13, Moscow, 119992 Russia}

\altaffiltext{2}{Special Astrophysical Observatory, Russian
Academy of Sciences, Nizhniy Arkhyz, Karachai-Cherkessian
Republic, 357147 Russia}

\begin{abstract}
We have used observations of the galaxy IC~10 at the 6-m telescope
of the Special Astrophysical Observatory with the SCORPIO focal
reducer in the Fabry--Perot interferometer mode and with the MPFS
spectrograph to study the structure and kinematics of ionized gas
in the central region of current intense star formation. Archive
VLA 21-cm observations are used to analyze the structure and
kinematics of neutral gas in this region. High-velocity wings of
the H$\alpha$ and [SII] emission lines were revealed in the inner
cavity of the nebula HL~111 and in other parts of the complex of
violent star formation. We have discovered local expanding
neutral-gas shells around the nebulae HL~111 and HL~106.
\end{abstract}


\maketitle

\section{INTRODUCTION}

The dwarf Irr galaxy IC~10 is the nearest galaxy with violent star
formation; it is often classified as a BCD galaxy based on its
high H$\alpha$ and IR~luminosity (Richer \textit{et al.} 2001).
The stellar population of IC~10 displays two bursts of star
formation; the first occured at least 350~million years ago, and
the second 4--10~million years ago (cf.~Hunter 2001; Zucker 2002;
Massey \textit{et al.} 2007; Vacca \textit{et al.} 2007, and
references therein). One consequence of the latest star-formation
burst is the multi-shell fine-filamentary structure of the
interstellar medium in IC~10: images of the galaxy in the
H$\alpha$ and [SII] lines reveal a giant complex of multiple
shells and super-shells, arc and ring structures, from 50~pc to
800--1000~pc in size (Zucker \textit{et al.} 2000; Wilcots and
Miller 1998; Gil de Paz \textit{et al.} 2003, Leroy \textit{et
al.} 2006; Chyzy \textit{et al.} 2003; Lozinskaya \textit{et al.}
2008, and references therein). The unusually high number of
Wolf-Rayet (WR) stars (whose spatial density in IC~10 is the
highest among dwarf galaxies and is comparable to that in massive
spiral galaxies) is a result of a recent star-formation burst that
was short but encompassed most of the galaxy (Massey \textit{et
al.} 1992; Richer \textit{et al.} 2001; Massey and Holmes 2002;
Crowther \textit{et al.} 2003; Massey \textit{et al.} 2007; Vacca
\textit{et al.} 2007, and references therein). This high density
of WR stars means that we are actually witnessing a short time
interval at the close of the latest  episode of star formation.

\begin{figure*}[t!]
\begin{center}
\includegraphics[scale=0.65]{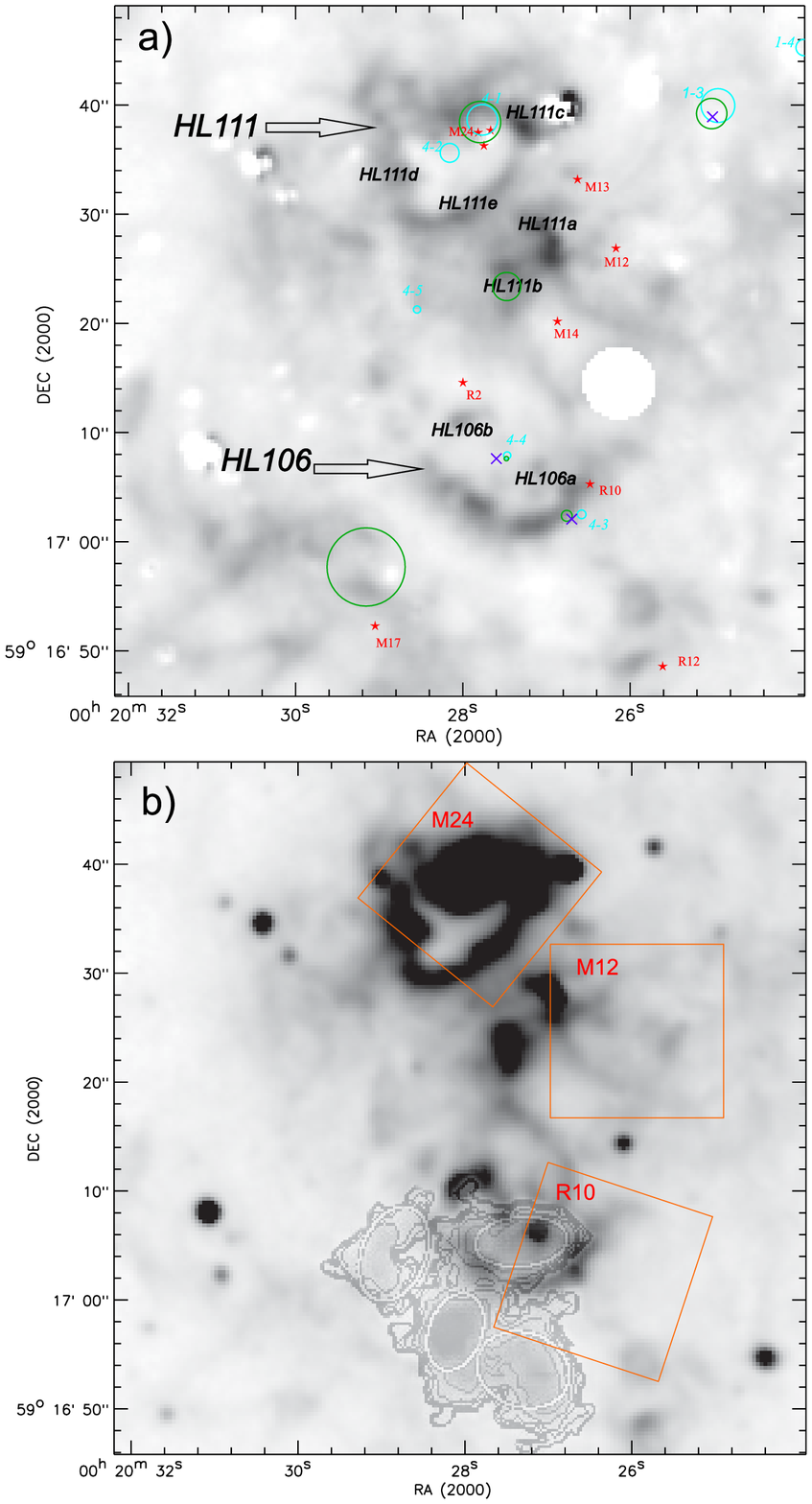}
\end{center}
\caption{(a): Image of the studied region of intense star
formation in the [SII] $\lambda(6717+6731)$~\AA\ lines (taken
through a narrow-band filter). Two shell nebulae, HL~111 and
HL~106, and their components are marked. The red asterisks show
WR~stars listed in Royer \textit{et al.} (2001) (``R'') or in
Massey and Holmes (2002) (``M''). Blue circles show star clusters
listed in Hunter (2001) (their names are shown as two digits);
green circles show  clusters listed in Tikhonov and Galazutdinova
(2009); blue crosses show  clusters listed in Sharina \textit{et
al.} (2010);  the diameters of the circles correspond to cluster
sizes. (b): H$\alpha$ image of the region (derived from FPI data),
overlaid with the intensity distribution (isophotes) of the
CO~emission according to Leroy \textit{et al.} (2006) (see the
text). Squares indicate the locations of the  MPFS fields labelled
according to their central WR stars. \hfill}
\end{figure*}

The central region of current vigorous star formation activity is
located in the galaxy's south-eastern sector. It  contains the
highest-density HI cloud, a CO molecular cloud, and a brightest
complex of emission nebulae about 300--400~pc in size, including
the two shell nebulae HL~111 and HL~106 (the names are from the
catalog Hodge and Lee (1990), as well as the youngest star
clusters and a dozen WR stars (Wilcots and Miller 1998; Gil de Paz
\textit{et al.} 2003; Leroy \textit{et al.} 2006; Lozinskaya
\textit{et al.} (2009), and references therein). Images of the
region in the H$\alpha$ and [SII] ($\lambda 6717+6731$~\AA) lines
are shown in Fig.~1.

According to Vacca \textit{et al.} (2007), the center of the
latest star formation episode is located near the object earlier
identified as the WR star M24. (Here and below, we use the prefix
``R'' for WR stars listed in Royer \textit{et al.} (2001) and
``M'' for those listed in Massey and Holmes 2002.) The nebula
HL~111c surrounding M24 is the brightest part of the shell HL~111
(Fig.~1); the size of the shell defined by its three arcs HL~111c,
HL~111d and HL~111e is about $10''$, or 35~pc for the distance
800~kpc. Hunter (2001) identified two clusters in the region of
HL~111: the brightest part, HL~111c, contains the cluster 4-1
which includes M24; the cluster 4-2 is in the middle of the
internal cavity. These are the youngest clusters in the galaxy,
with ages of three to four million years~(Hunter 2001).

It was noted earlier that the WR~star M24 actually consists of the
three blue stars M24-A, M24-B, and M-24C, separated by
$1''{-}2''$; one of the components was suspected to be an O3If
star, a WN~star, or a cluster containing WN~stars~(Crowther
\textit{et al.} 2003). Vacca \textit{et al.} (2007) demonstrated
that M24 was actually a close stellar group consisting of at
least of six blue stars (three blue stars were identified in M24A
and two in M24B). Four of these six stars are WR~candidates.

The southern part of the above-mentioned HI and CO~cloud -- the
highest-density one in the galaxy -- contains the shell nebula
HL~106, whose sources of ionizing radiation are probably the WR
stars R2 and R10, as well as the clusters 4-3 and 4-4 (Fig.~1).
According to the estimates by Hunter (2001), the clusters are
appreciably older than the young clusters 4-1 and 4-2 in the
HL~111 region.

The aim of the current paper is a detailed study of the structure
and kinematics of the ionized and neutral gas in the region of
violent star formation. Our study is based on observations with
the 6-m telescope of the Special Astrophysical Observatory (SAO)
using the SCORPIO focal reducer working in the Fabry--Perot
interferometer (FPI) mode. We also use observations with the MPFS
panoramic spectrograph and with SCORPIO in the long-slit
spectroscopy mode (the latter observations are discussed in
detail in Lozinskaya \textit{et al.}, 2009). The following
sections describe our observations, and present and discuss our
results, while the last section summarizes our conclusions.

\begin{table*}[t!]

 \caption{The log
of observations}
\begin{center}
\begin{tabular}{|l|c|c|c|c|c|c|}
\hline \multicolumn{1}{|c|}{Mode} &   \multicolumn{1}{c|}{Field}
& Date        &  \multicolumn{1}{c|}{Range}  &
$\delta\lambda$,~\AA & $T_{exp}$, s & \parbox[c][1cm]{2.2cm}{Seeing}\\
\hline
FPI & The whole galaxy & $2005.09.08/09$  &  H$_{\alpha}$        & 0.8  &$10\;800$ & $0.8''{-}1.3''$ \\
FPI & The whole galaxy & $2008.10.23/24$  &  [SII] $\lambda 6717$& 0.8  &$12\;960$ & 1.4--2.0 \\
\hline
MPFS & M24 & 2004.08.08/09& 3990--6940 & 6.5 & 2700 & 1.7--2.0 \\
MPFS & R10 & 2004.08.08/09& 3990--6940 & 6.5 & 1800 & 1.4 \\
MPFS & M12 & 2005.09.28/29& 3990--6940 & 6.5 & 3600 & 2.0 \\
 \hline
\end{tabular}
\end{center}
\end{table*}

All radial velocities in this paper are heliocentric. The
distance to the galaxy is assumed to be 800~kpc (the angular
scale is ${\simeq}3.6$~pc per arcsecond; cf. Section~4).

\section{OBSERVATIONS AND DATA REDUCTION}

A log of our observations with the SAO 6-m telescope using the
SCORPIO focal reducer in the FPI mode and the MPFS is presented
in the Table. Its columns contain (1) the instrument, (2) the
field covered or the designation of the MPFS field (according to
the central WR star), (3) the observation epoch, (4) the
wavelength range, (5) the spectral resolution, (6) the exposure
time, and (7) the seeing.

\subsection{Observations with the Fabry--Perot Interferometer}

Our observations with the 6-m telescope using SCORPIO and the
scanning FPI were made in two lines: H$\alpha$ and [SII]
$\lambda6717$~\AA. The capabilities of the SCORPIO for
interferometric observations are described in detail by Moiseev
(2002). We used a scanning FPI working in the 501st interference
order at $\lambda6562.78$~\AA. In these observations, the
distance between adjacent interference orders, $\Delta
\lambda=13$~\AA\ for H$\alpha$ line and $\Delta\lambda=13.7$\AA\
for the [SII] $\lambda6717$~\AA\ line, corresponded to a region
free of overlapping orders ${\sim}600$~km/s wide in radial
velocity. The width (FWHM) of the instrumental profile was about
$0.8$~\AA, ${\sim}35$~km/s. We performed preliminary
monochromatization using interference filters with a FWHM of
${\sim}15$~\AA, centered on the H$\alpha$ and [SII]
$\lambda6717$~\AA\ lines. The detector was a
$2048\times2048$-pixel EEV 42-40 CCD array. To reduce the readout
time, we observed with $2\times2$-pixel binning in the H$\alpha$ line
and $4\times4$-pixel binning in the [SII] $ \lambda6717$~\AA\
line. In this way, we obtained $1024\times1024$-pixel images for
H$\alpha$ and $512\times512$-pixel images for the [SII] line in
each spectral channel, with the scales being $0.36''$ per pixel
and $0.71''$ per pixel, respectively; the full field of view was
$6.1'$.

In the course of our observations, we obtained a sequence of
36~interferometric images of the object for different distances
between the FPI plates, so that the width of a spectral channel
was $0.36$~\AA\ in H$\alpha$ and $0.38$~\AA\ in the [SII] $
\lambda6717$~\AA\ line, i.e. about 17~km/s in radial velocity. To
correctly remove stray ghost images from numerous regions of
emission in the galaxy, we observed at two different alignments
of the instrument's field of view. The total exposure time was
10800~s for observations in H$\alpha$ and 12960~s for the [SII]
$\lambda6717$~\AA\ line.

We reduced the observations using IDL-based software (Moiseev
2008). After the preliminary data reduction, the data were
presented as data cubes with size of $1024\times1024\times36$
and  $512\times512\times36$, where an each spatial element
(``spaxel'') contains a 36-channel spectrum. We removed ghost
images using the algorithms described by Moiseev and Egorov
(2008). Each time, we observed for two orientations of the field
of view and then compiled the resulting preliminary data cubes.
This procedure guarantees the removal of ghosts due to bright
stars, which is important for our studies of spectral-line
shapes. The final angular resolutions (after averaging during the
data reduction) were $1.3''$ and $2.2''$ in H$\alpha$ and [SII]
$\lambda6717$~\AA\ lines, respectively. The uncertainty in the
wavelength scale was 3-5 km/s.

Using narrow-band filters for preliminary mono\-chro\-matization
of the light considerably reduced the flux from neighboring
spectral lines in overlapped interference orders. A possible
residual contribution from the second component of the [SII]
$\lambda6731$~\AA, when observing in the [SII] $\lambda6717$~\AA\
line may be present near ${-}300$~km/s, with the system velocity
being ${-}330$~km/s (thus, at ${+}30$~km~/s relative to the
maximum of the primary line component). There may be
contributions from the [NII] $\lambda6584$~\AA\ line to H$\alpha$
at about ${-}587$~km/s and ${+}12$~km~/s (${-}260$ and
${+}340$~km/s from the principal component); from the [NII]
$\lambda6548$~\AA\ line at about ${-}405$~km/s (${-}75$~km/s
relative to the main component); and the $\lambda6553.62$~\AA\
night sky line, if it is not completely subtracted, at
${-}420$~km/s.

\subsection{MPFS Observations}

We observed selected regions in the galaxy using the
integral-field Multi-Pupil Fiber Spectrograph (MPFS) at the
primary focus of the 6-m telescope (Afanasiev \textit{et al.}
2001; see also http://www.sao.ru/hq/lsfvo/devices.html). The
detector was a $2048\times2048$-pixel EEV 42-20 CCD array. The
spectrograph enables simultaneous recording of spectra from 256
spatial elements (structurally designed as square lenses) that
form an array of $16\times16$ elements in the plane of the sky.
The angular size of each image element is $1''$.

In this study, we partially use spectra for three of the six
fields observed for IC10 with a resolution of about 6.5~\AA\ in
the range $3990{-}6940$~\AA, their positions are shown in Fig.~1b.

We reduced our observations using software developed at the SAO
Laboratory of Extragalactic Spectroscopy and Photometry, working
in the IDL environment. The result of this reduction is a data
cube in which each $16''\times16''$ image element is represented
by a spectrum of 2048 elements.

We also used images of the galaxy in the [SII] and H$\alpha$
lines that we obtained earlier (Lozinskaya \textit{et al.} 2008).
The image in the [SII]~$\lambda$(6717+6731)~\AA\ lines displayed
in Fig.~1a is based on narrow-band direct imaging with SCORPIO;
the monochromatic H$\alpha$ image in Fig.~1b was obtained by
integrating the fluxes in the spectra derived from FPI
observations.

\section{RESULTS OF THE OBSERVATIONS}

\subsection{Kinematics of Ionized Gas}

We studied the kinematics of the ionized gas in the region of
violent star formation, including HL 111 and HL 106, using the
observations with the scanning FPI in the H$\alpha$ and [SII]
$\lambda6717$~\AA\ lines. We also used the MPFS observations for
comparison.

\begin{figure*}[p!]
\begin{center}
\includegraphics[scale=0.85]{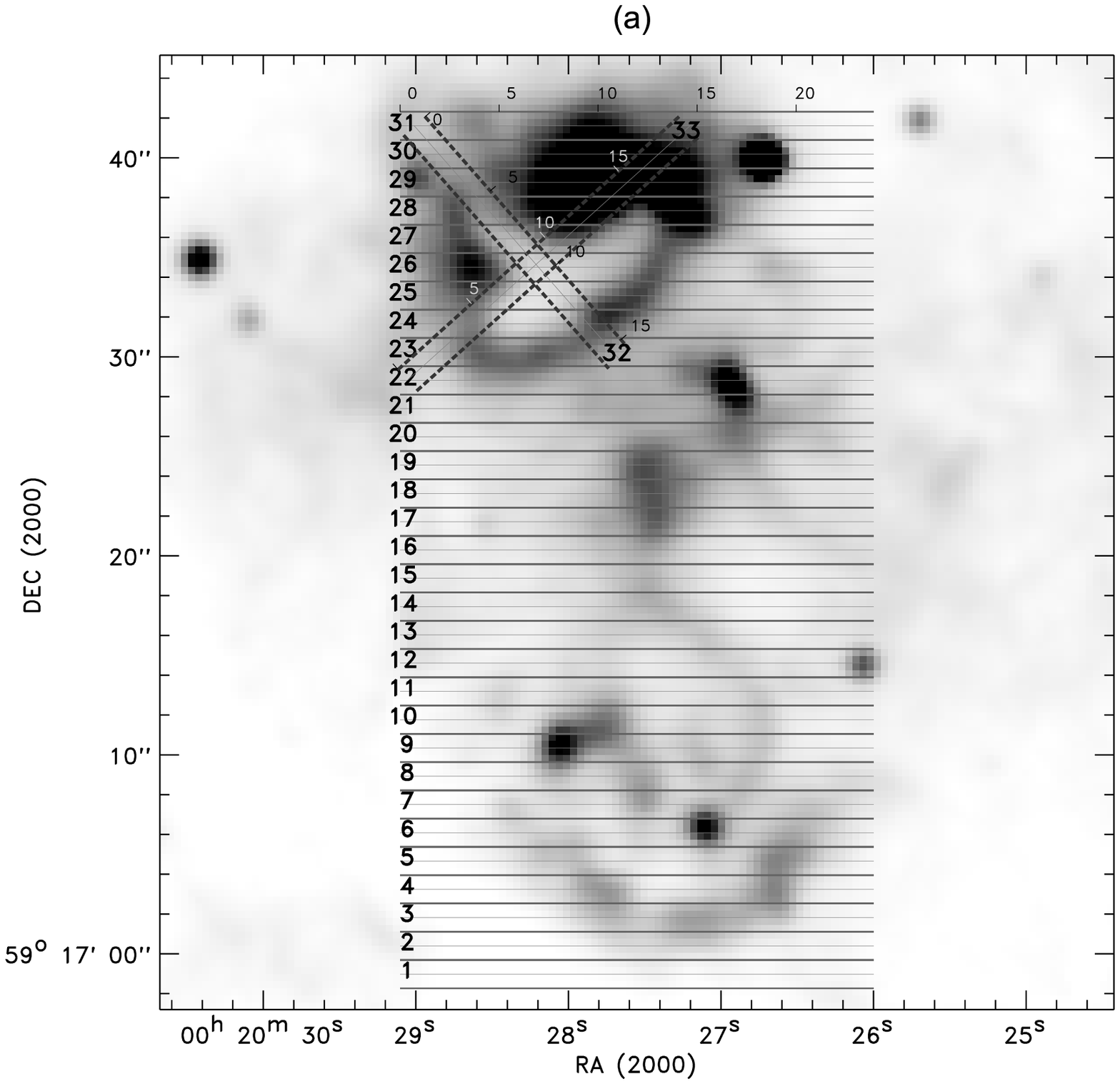}
\end{center}
\vspace*{-5pt} \caption{(a) Positions of the sections along which
we plotted $P/V$ diagrams in the H$\alpha$ and [SII]
$\lambda6717$~\AA\ lines derived from the FPI observations. The
numbers top of this map and all subsequent figures are the
``markings'' of the sections in arcseconds; the markings for two
sections, No.~32 and No.~33, are indicated directly on them. (b)
Sample $P/V$ diagrams for H$\alpha$. (c) Sample $P/V$ diagrams for
the [SII] $\lambda6717$~\AA\ line. The arrows in panels (b) and
(c) indicate the regions whose the H$\alpha$ and
[SII]$\lambda6717$\AA\ profiles are displayed in Fig.~4. \hfill}
\end{figure*}

\begin{figure*}[p!]
\begin{center}
\includegraphics[scale=0.9]{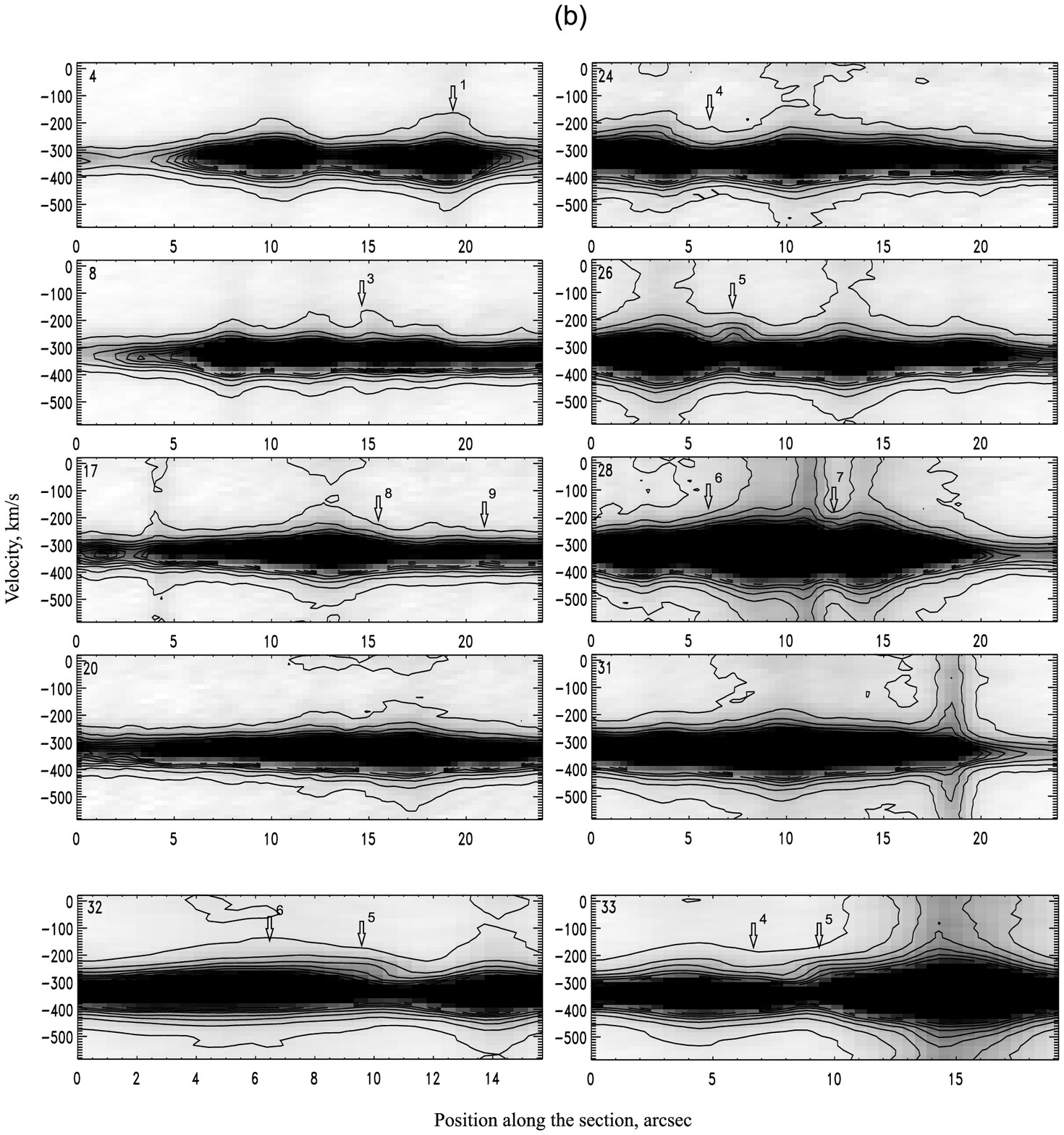}
\end{center}
\addtocounter{figure}{-1} \vspace*{5pt} \caption{(Contd.) \hfill}
\end{figure*}

\begin{figure*}[p!]
\begin{center}
\includegraphics[scale=0.9]{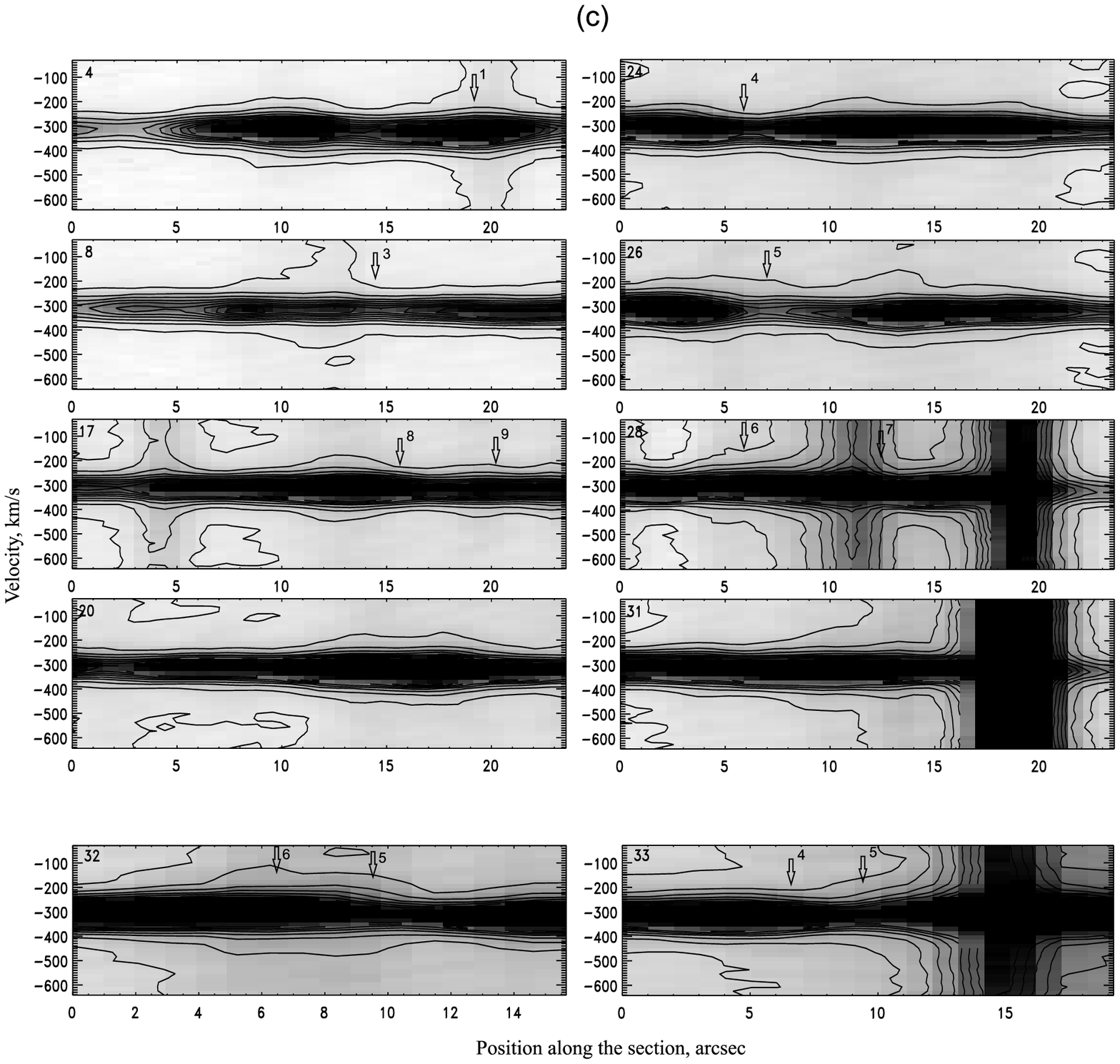}
\end{center}
\addtocounter{figure}{-1} \vspace*{5pt} \caption{(Contd.) \hfill}
\end{figure*}

\begin{figure*}[t!]
\begin{center}
\includegraphics[scale=0.75]{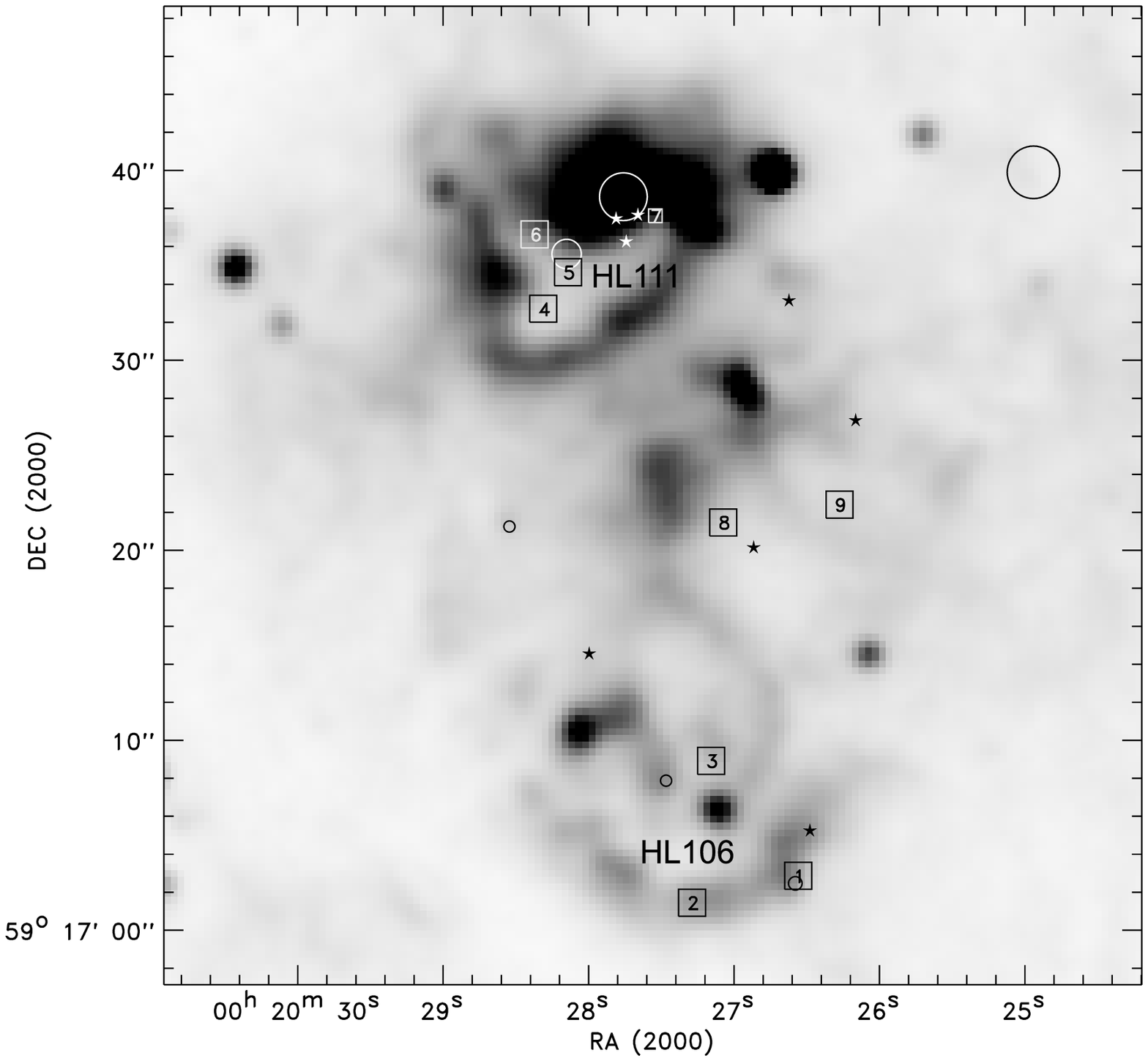}
\end{center}
\vspace*{5pt} \caption{Positions of the H$\alpha$ and [SII]
$\lambda6717$~\AA\ profiles presented in Fig. 4 below, shown by
squares with numbers inside, which correspond to the area
numbers. As in Fig. 1, circles and asterisks denote star clusters
and WR~stars. (The parts of the $P/V$ diagrams used to plot the
profiles are marked with asterisks in Figs.~2b and 2c.) \hfill}
\end{figure*}

\begin{figure*}[p!]
\begin{center}
\includegraphics[scale=0.65]{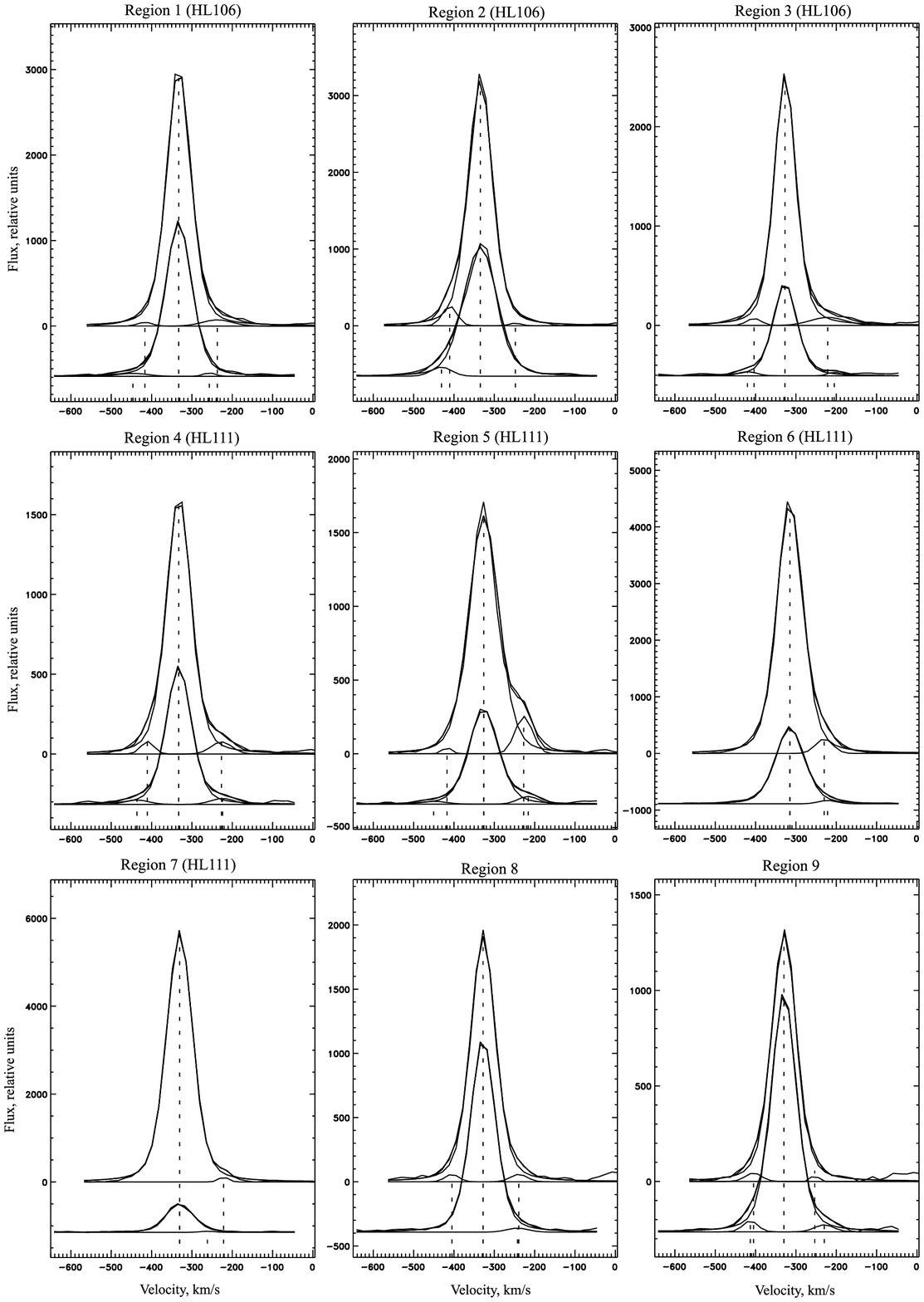}
\end{center}
\vspace*{10pt} \caption{Sample line profiles for H$\alpha$
(upper) and [SII] $\lambda6717$\AA\ (lower) from the FPI
observations (in relative units). Plotted are the observed
profiles, a fit of the main components with Voigt profiles, the
high-velocity components identified as excesses relative to the
Voigt wings (fit with Gaussians), and the sums of the identified
components (which virtually coincide with the observed profiles).
The region numbers indicated above correspond to the numbers in
the squares in Fig.~3. \hfill}
\end{figure*}

Taking into account the complex, multi-shell, fine-filamentary
structure of the studied region and the presence of the multiple
sources of stellar wind mentioned in the Introduction, the region
required an extremely detailed analysis of its radial-velocity
field.

The first stage of the analysis consisted of a systematic search
for areas with high (supersonic?) velocities or other kinematic
peculiarities. For this purpose, we used our scanning-FPI
observations to cover the entire program region in two
perpendicular directions with a grid of abutting sections, along
which we plotted diagrams of the gas radial-velocity distribution
(so-called $P/V$ diagrams). We performed a search of high-velocity
features   in both H$\alpha$ and [SII] $\lambda6717$~\AA\ lines
observations.

Figure~2 presents an image in the H$\alpha$ line showing the
positions of the sections in the east-west direction. We plotted
$P/V$ diagrams in both lines along these sections. We show the
most interesting diagrams in the H$\alpha$ and [SII]
$\lambda6717$~\AA\ lines as examples. To convenience, the
sections in the figure have ``markings'' in arcseconds. The
arrows in Figs.~2b, 2c indicate the places used to plot the
profiles of the two lines shown in Fig.~4 below.

The second stage of our study consisted in a more detailed
analysis of the kinematics for those regions where our systematic
search had revealed velocity shifted line features. For this
purpose, we plotted $P/V$ diagrams for individual nebulae on a
large spatial scale and also considered the profiles of both
lines.

Two $P/V$ diagrams of this kind and their positions on the image
of the nebula HL 111 are shown in Fig.~2 (sections No.~32 and
No.~33). These diagrams clearly show weak high-velocity features
inside the cavity (marked with arrows).

The positions of the H$\alpha$ and [SII] $\lambda6717$~\AA\
profiles obtained from our FPI observations are shown in Fig.~3.
Figure~4 presents the corresponding H$\alpha$ (top) and [SII]
(bottom) profiles in relative units.

When deriving the profiles, we fit the main component of the line
with a Voigt function. Weak high-velocity components were
identified as excesses above the Voigt wings; the residual line
features remaining after subtraction of the Voigt profile were
fitted with Gaussians. Thus, the figure displays for each line:
the observed profile, its fit with a Voigt function, its
high-velocity components identified as excesses above the Voigt
wings (fitted with Gaussians), and a sum of all the identified
components (which virtually coincides with the observed profile).

Summarizing the results of our analysis of the ionized-gas
kinematics in the region of violent star formation, we can draw
the following conclusions.

According to our FPI observations, the gas velocity in the region
containing the nebulae HL 111 and HL 106 is from ${-}315$~km/s to
${-}355$~km/s. Velocity variations from ${-}318$ to ${-}335$~km/s
can be noted in the [SII] line. The mean velocity of the ionized
gas in the region of violent star formation is ${-}330$~km/s.

These findings agree with the H$\alpha$ measurements of Thurow and
Wilcots (2005) (whose observations cover one-third of the bright
star-forming region we are studying, in its western part; see
Fig.~5 in Thurow and Wilcots 2005). The neutral gas in the region
has the same mean velocity (Willcots and Miller 1998; see below).

The $P/V$~diagrams plotted from our FPI observations in H$\alpha$
and the [SII]~lines did not show a classical ``velocity ellipse''
(with the radial velocities of the front and back sides of the
expanding shell decreasing in magnitude with radius in the plane
of the sky). The individual cross sections presented in Fig.~2
reveal slight deviations of the velocity of the line maximum from
the mean velocity derived for ``unperturbed'' gas in the region,
but such deviations never exceed 15–-20 km/s. The velocities
determined for the line maximum, i.e. those corresponding to the
brightest emission, suggest a possible expansion velocity of the
HL~111 shell within 10–-15 km/s. At the same time, some regions of
the nebula HL~111 exhibit profile asymmetry or weak features in
the red line wing, at a level of about $2\%{-}6\%$ of the maximum
intensity, at velocities between ${-}280$~km/s and ${-}240$~km/s.
High-velocity features in the blue line wing are seen between
${-}450$~km/s and ${-}410$~km/s.

It follows from Figs.~3 and 4 that the most prominent
high-velocity features are observed in the red wing of the
H$\alpha$ line, in an irregularly shaped cavity inside HL~111.

A similar pattern is observed for the nebula HL~106. The
velocities measured for the maximum of the H$\alpha$ line and
characterizing the brightest clumps in the shell exhibit small
variations, and imply a possible mean expansion velocity not
exceeding 10-–15~km/s. Weak features in the H$\alpha$ wings, at
about $2\%{-}10\%$ of the maximum intensity, are observed in
HL~106 at velocities between ${-}260$~km/s and ${-}250$~km/s and
between ${-}450$~km/s and ${-}420$~km/s.

There also exist appreciable features in the red wing of the
[SII] $\lambda6717$~\AA\ line in the inner cavity of the nebula
HL 111, at the same velocities as for H$\alpha$ (Fig.~4). The good
coincidence between the velocities of weak features in the wings
of both lines is also observed in other regions of the studied
complex, supporting their reality. The fact that some regions of
the complex display weak H$\alpha$ wings, while the [SII]
$\lambda6717$~\AA\ observations do not indicate such wings, can
be explained by the low intensity of the latter line (it is about
an order of magnitude weaker than H$\alpha$).

We used MPFS observations of three fields in the region to check
the FPI results. The MPFS observations were used to derive
average profiles of the H$\alpha$, [SII] $\lambda6717$~\AA, and
[NII] $\lambda6584$~\AA\ lines for the three fields shown in
Fig.~1b.

We do not reproduce the corresponding line profiles here because
the MPFS observations have a much poorer spectral resolution than
the FPI observations. However, our MPFS observations fully confirm
the presence of the weak H$\alpha$ and [SII] $\lambda6717$~\AA\
wings detected with the FPI. Due to their low spectral
resolution, the velocities of the weak line components sometimes
differ considerably from those derived from the FPI observations
(by as much as 50--100~km/s). Therefore, we used the MPFS data
only to verify the presence of red line wings in these regions,
while all our velocity estimates were made from the FPI
observations.

We wish to emphasize that weak features can be detected in line
wings in several regions of this complex of violent star
formation, but their intensities do not exceed $2\%{-}10\%$ of
the line peak intensity. The [SII] $\lambda6717$~\AA\ profile
appears more symmetric, but this line's integrated brightness is
between $4\%$ and $12\%$ of that of the H$\alpha$ line, making
the identification of weak wings less reliable in this case.

The most prominent high-velocity features in the red H$\alpha$
wings are observed in an irregularly shaped cavity inside the
shell of HL~111, near the cluster 4-2.

Thurow and Wilcots (2005) noted the presence of red H$\alpha$
wings in the western part of the region between HL~111 and HL~106.
However, the velocities of the feature left after subtraction of
the main line component, given in Table~4 of Thurow and Wilcots
(2005) (${-}282 \div {-}256$~km/s), differ from our measurements
for the same places.

\subsection{Kinematics of Neutral Gas}

The large-scale kinematics of the neutral hydrogen in IC~10 was
studied by Wilcots and Miller (1998). The HI~velocity dispersion
in the complex of violent star formation and, in particular, in
the extended surroundings of M24 is not the highest in the galaxy
(Wilcots and Miller 1998, Fig.~14). The mean velocities at the
maximum of the 21-cm line give a possible expansion velocity of
the large arc structures revealed by Wilcots and Miller (1998) not
exceeding 10--15~km/s.

We reanalyzed the data cube of VLA observations of IC~10 in the
21~cm line, which was made available to us, to study in more
detail the ``local'' structure and kinematics of the HI in the
close neighborhood of the star-forming complex and the brightest
nebulae, HL~111 and HL~106. For this purpose, we used data with
an angular resolution of $4.7''\times 5''$, or about 20~pc for the
distance of 800~kpc.

The results of our analysis are presented in Figs.~5 and~6.
Figure~5a shows the H$\alpha$ image of the bright region of
violent star formation, overlaid with the integrated brightness
distribution in the 21~cm line. Figure~5b shows the same H$\alpha$
image overlaid with the distribution of neutral hydrogen emitting
at ${-}336$~km/s, close to the velocity of ionized gas in HL~111
and HL~106. Figure~5a indicates the presence of a band with the
highest column density of neutral hydrogen, extending from
northeast to southwest between HL~106 and the Synchrotron
Superbubble that coincides with the absorbing dust band in the
optical images of the region. (The unique Synchrotron Superbubble
in IC~10, earlier believed to be formed by multiple explosions of
about a dozen supernovae (Yang and Skillman 1993; Thurow and
Wilcots 2005), was demonstrated by Lozinskaya and Moiseev (2007)
to be a hypernova remnant.)

\begin{figure*}[t!]
\begin{center}
\includegraphics[scale=0.65]{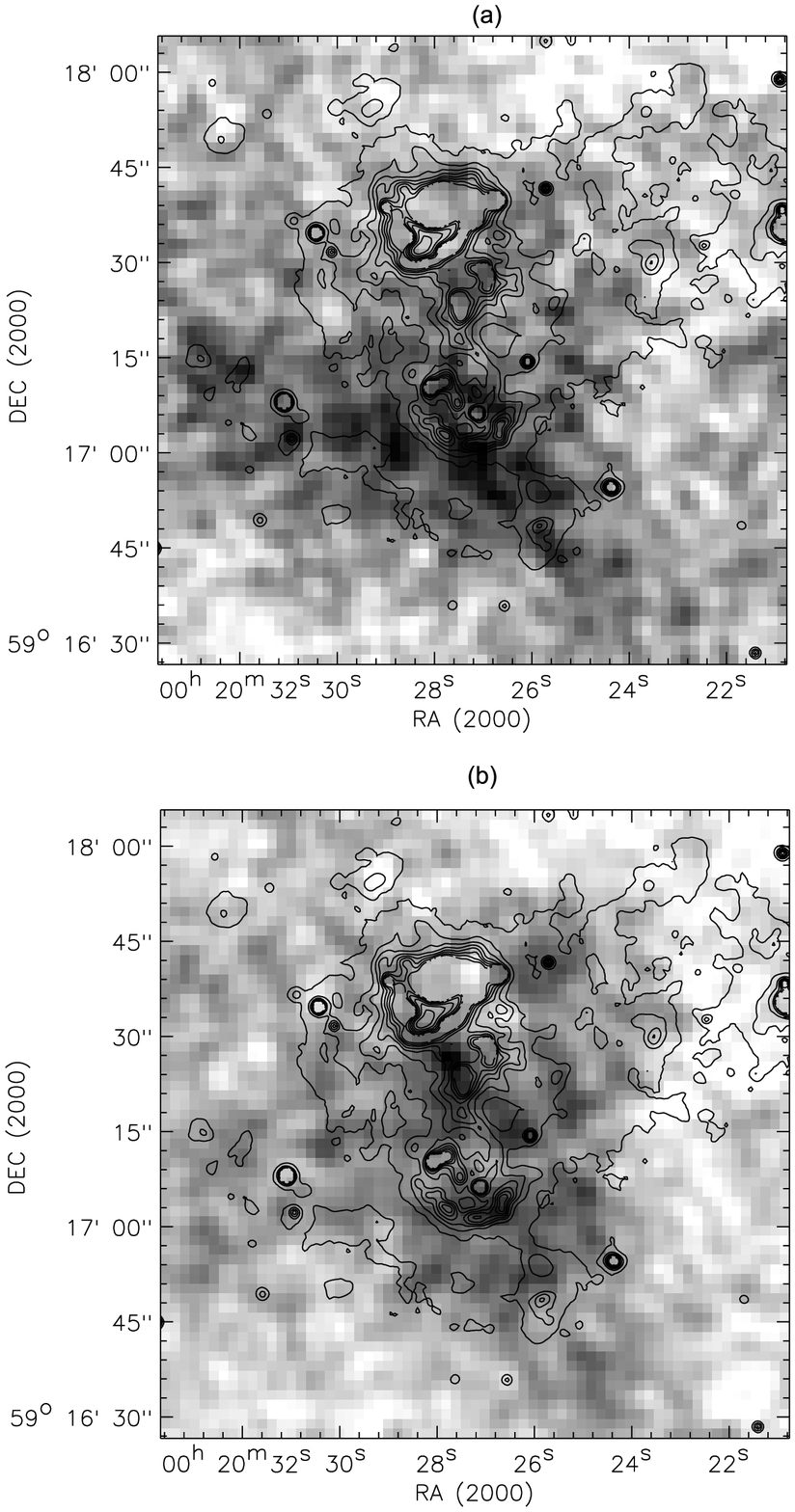}
\end{center}
\caption{(a) H$\alpha$ image of the bright star-forming region
(plotted as isophotes) overlaid with the distribution of the
integrated brightness in the region in the 21~cm line. (b)
H$\alpha$ image overlaid with the distribution of neutral hydrogen
emitting at a velocity of ${-}336$~km~s$^{-1}$. \hfill}
\end{figure*}

\begin{figure*}[p!]
\begin{center}
\includegraphics[scale=0.65]{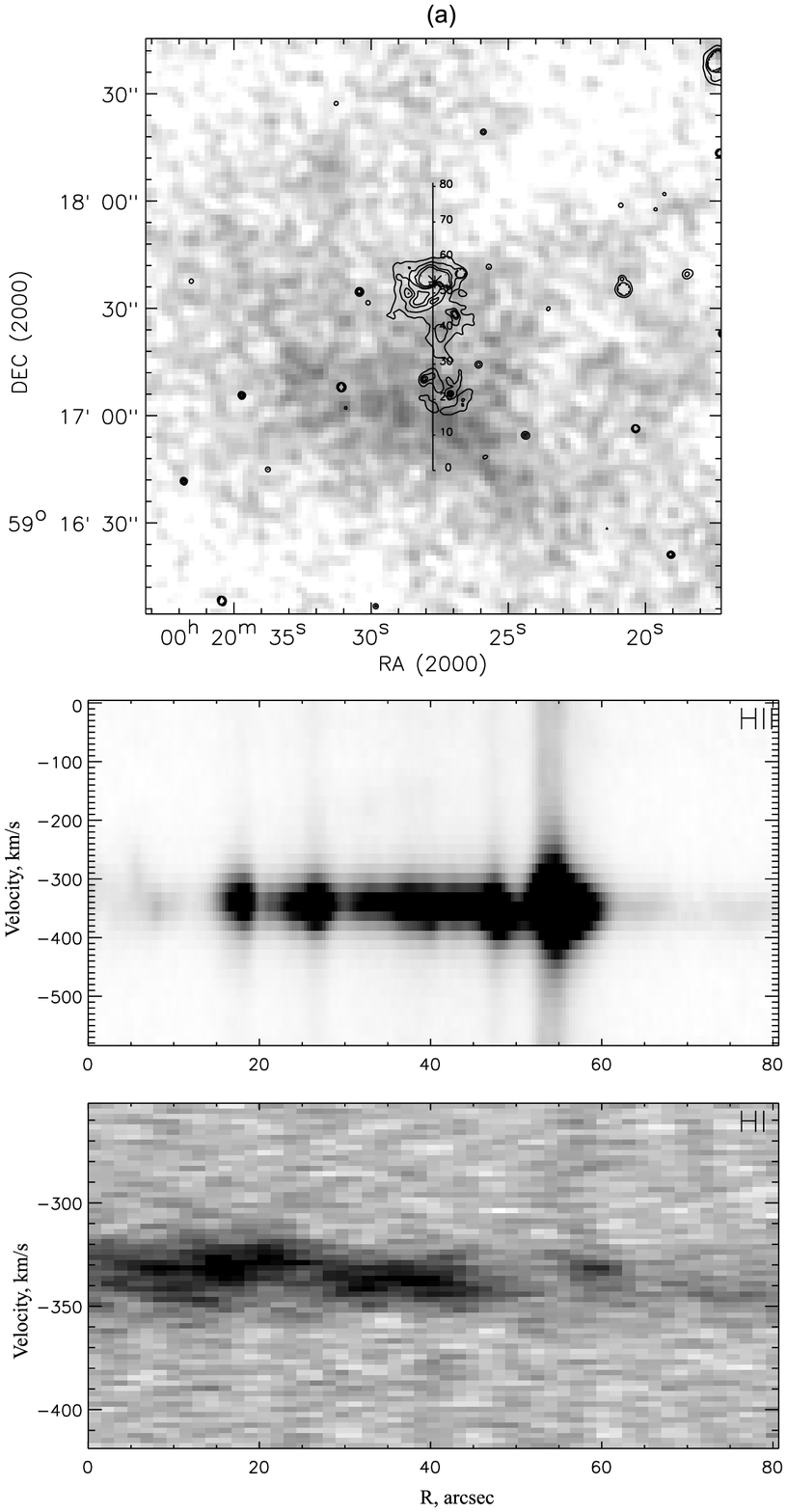}
\end{center}
\vspace*{3pt} \caption{$P/V$~diagrams plotted for emission of
ionized and neutral gas in the same direction. The three upper
maps show the positions of the sections in the overlaid images of
the region in the 21~cm and H$\alpha$ (isophotes) lines; the
three panels in the middle -- the corresponding $P/V$~diagrams in
the H$\alpha$ line, and the three bottom panels -- the
$P/V$~diagrams in the 21~cm line. \hfill}
\end{figure*}

\begin{figure*}[t!]
\begin{center}
\includegraphics[scale=0.65]{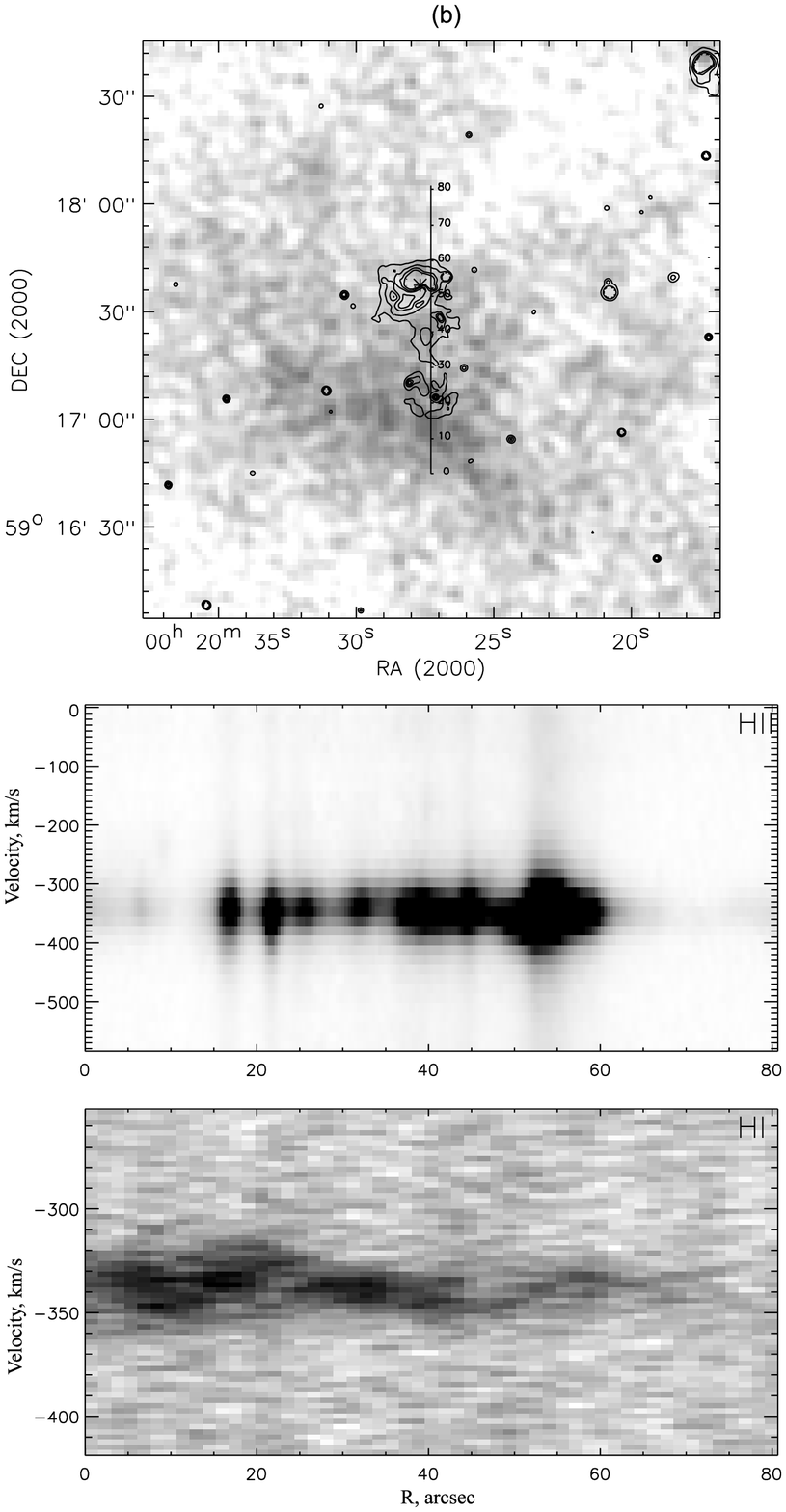}
\end{center}
\addtocounter{figure}{-1} \vspace*{5pt} \caption{(Contd.) \hfill}
\end{figure*}

\begin{figure*}[t!]
\begin{center}
\includegraphics[scale=0.65]{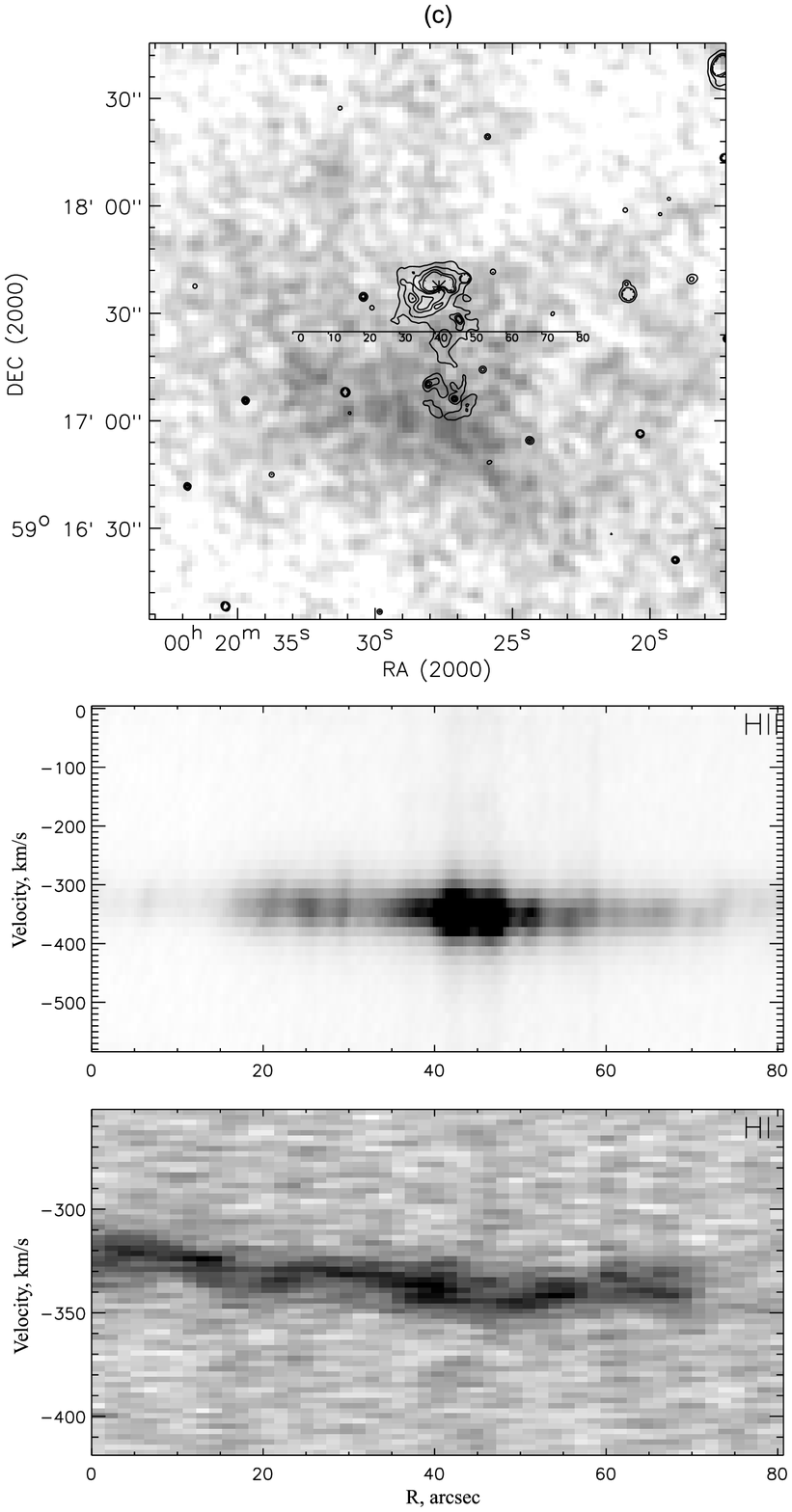}
\end{center}
\addtocounter{figure}{-1} \caption{(Contd.) \hfill}
\end{figure*}

It follows from Fig.~5b that a ``local'' extended HI~cloud
emitting at the same velocity as the ionized gas is observed in
the vicinity of the bright star-forming region that includes
HL~111 and HL~106.

Figures~6a--6c present some of our $P/V$ diagrams for the emission
of the ionized and neutral gas in the same direction. The
positions of the corresponding directions are shown in the
overlaid images of the region in the 21 cm and H$\alpha$ lines
(isophotes).

 It follows from the displayed $P/V$~diagrams that an
expanding shell of neutral gas is present just outside the ionized
shell nebula HL~111 (see the positions 37--57 in Fig.~6a, 35--56
in Fig.~6b, 35--56 in Fig.~6c). The mean radial velocity of this
outer neutral shell is ${-}333 \div {-}336$~km/s, i.e. it
coincides with the mean velocity of the optical nebula. The mean
expansion velocity of the HI shell surrounding HL~111 is
$10{-}15$~km/s, so that the receding side of the neutral shell has
a mean velocity of about 10~km/s and the approaching side a mean
velocity of about 15~km/s. Weaker features of the expanding
neutral gas shell are observed at velocities differing by up to
${\simeq} 30$~km/s from the mean ``unperturbed'' velocity.

Thus, apart from the extended shells of neutral gas, about
$1'{-}1.5'$ in size visible in the entire field of the galaxy in
Fig.~9 of Wilcots and Miller (1998), our detailed analysis of
21~cm observations has revealed a neutral shell about
$25''{-}30''$ in size around the optical nebula HL~111. The mean
expansion velocity of this shell is $10{-}15$~km/s; weak features
of the 21~cm line correspond to an expansion velocity about
30~km/s. This expanding neutral shell could have formed under the
influence of wind from a group of WR~stars that belong to M24 as
well as from the above-mentioned clusters 4--1 and 4--2 (cf.
Section~4).

It follows from Fig.~6 that a similar ``local'' shell, expanding
at a mean rate of $15 {-} 20$~km/s, is also observed around the
nebula HL~106 in the southern part of the region of violent star
formation (positions 4--29 in Fig.~6a, 4--28 in Fig.~6b, 8--27 in
Fig.~6c). Here, we also observe weaker features corresponding to
expansion at velocities up to 25~km/s. The sources of mechanical
energy responsible for the formation of the HI shell around
HL~106 are probably the WR~stars R2 and R10 and the clusters 4--4
and 4--3.

Note that both the detected neutral shells have non-uniform
brightness distributions and a clumpy structure. The receding
side of the HI shell around HL~111 is fainter than the
approaching side, in virtually all sections crossing it in
different direction. In contrast, as a rule, the receding side of
the neutral shell around HL~106 is brighter.

The third link in the chain of expanding shells situated in the
high-density cloud of neutral gas is the HI shell around the
Synchrotron Superbubble (Lozinskaya \textit{et al.} 2008). This
third shell is at the opposite side of the dense gas and dust
band and expands with the highest mean velocity of $25{-}30$~km/s,
which can be naturally explained with a more powerful energy
source, such as a hypernova explosion (Lozinskaya and Moiseev
2007).

The cross-section shown in Fig.~6c passes the faint regions of
ionized gas between HL~111 and HL~106. We also see local
``perturbations'' of neutral gas here, against the smooth change
of the mean HI velocity in the direction from east to west. The
highest neutral-gas velocities vary between ${\sim}30$~km/s and
${\sim}{-}30$~km/s relative to the mean velocities for the
corresponding regions. Similar local neutral-gas
``perturbations'' are also visible in other $P/V$~diagrams, to the
east and west from the bright region of violent star formation (to
save space, we do not reproduce them here).

\section{DISCUSSION}

The galaxy IC~10 is a very promising object for studies of the
interaction of stellar winds and supernovae with interstellar gas
in a region of intensive star formation. First, it is the nearest
Irr galaxy with violent star formation; second, it follows from
the arguments presented in the Introduction that the most recent
star formation event in the galaxy is also the closest overall in
time. We are studying the structure and kinematics of ionized and
neutral gas in the region of the recent star formation burst in
detail. This bright region includes the nebulae HL~111 and HL~106
and is located in the northern part of the highest density HI
cloud in the galaxy (Wilcots and Miller 1998), connected with the
highest-density cloud of molecular gas (Leroy \textit{et al.}
2006; Bolatto \textit{et al.} 2000) and the brightest IR~emission
from heated dust (Bolatto \textit{et al.} 2000). The column
density of this cloud is $N\textrm{(HI)} > 2.5\times
10^{21}$~cm$^{-2}$; according to Wilcots and Miller (1998) and
Leroy \textit{et al.} (2006), the column density towards M24 is
$N({\textrm{HI}}+{\textrm{H}}_{2})= 2.2\times 10^{21}$~cm$^{-2}$.
This is above the star formation threshold for dwarf galaxies
($N({\textrm{HI}})> (0.4-1.7)\times 10^{21}$~cm$^{-2}$) according
to Begum \textit{et al.} (2006), and can naturally explain the
location of the center of the latest violent star-formation event
at this position in IC~10.

Early {\textbf{estimates of light extinction and distance to
IC~10}} varied strongly due to the low galactic latitude of IC~10;
current estimates agree.

According to our estimates (Lozinskaya \textit{et al.} 2009), the
color excess for the galaxy's HII regions is $E(B-V)=0.8^{m}\div
1.1^{m}$; based on old stellar populations, Vacca \textit{et al.}
(2007) derived $E(B-V)=0.95^{m}$, $(m-M)_{0}=24.48^{m}\pm
0.08^{m}$. A similar value, $(m-M)=24.30^{+0.18}_{-0.10}$, was
found by Kniazev \textit{et al.} (2007) from planetary nebulae;
Sanna \textit{et al.} (2008a,b) used three different methods to
derive the distance moduli $24.51^m\pm 0.08^{m}$, $24.56^m\pm
0.08^{m}$, and $24.60^m\pm 0.15^{m}$.  According to Tikhonov and
Galazutdinova (2009) $(m-M)=24.47^{m}$, $d=780\pm 40$~kpc.  Taking
into account all these estimates, we take the distance to IC~10 to
be $d=800$~kpc for the purposes of this study. (The  recent
estimate by Kim \textit{et al.} (2009), $d=715$~kpc, does not
significantly change our results.)

The color excess in the region of HL~111 is somewhat below the
mean level for the galaxy derived from old stars and below the
reddening estimated for other HII regions (Hunter 2001; Vacca
\textit{et al.} 2007; Lozinskaya \textit{et al.} 2009, and
references therein).

Our long-slit spectrograph observations yielded
$E(B-V)=0.54^m\pm0.15^{m}$ in the nebula HL~111 and $E(B-V) =
1.10^m\pm0.15^{m}$ in HL~106. From the MPFS observations in this
study (the M24 field in Fig.~1b), we derived the mean
$E(B-V)=0.84^m\pm0.14^m$ in the bright region of HL~111 and
$E(B-V)= 0.73^m\pm0.12^m$ inside the cavity. We can identify two
local maxima here: the color excess in the north-west of the
central bright part, HL~111c, reaches $0.9^m$ and in the
south-east, $1.0^m$.

Vacca \textit{et al.} (2007) obtained based on young blue stars
$E(B-V)=0.6^{m}$ for the region of HL~111, in agreement with our
estimates. In the central part of the galaxy closest to HL~111
(field S4), the color excess $E(B-V)= 0.60^m\pm 0.14^m$ was
derived by Sanna \textit{et al.} (2008b), and a reddening decrease
from the central star-forming region to the periphery of the
galaxy was detected. Both these findings are easy to explain. On
the one hand, the central star-forming region is adjacent to the
highest-density cloud of neutral and molecular gas in the galaxy,
and it is quite probable that some of its regions are embedded in
this cloud. On the other hand, a local decrease in the reddening
could mean that the wind from young stars of the most recent
star-formation event has swept away the surrounding gas.

{\textbf{The HL~111 shell around M24}} is the brightest and most
interesting object in the star-forming complex. As is noted in
the Introduction, the brightest star~M24, which was earlier
resolved into three components, is actually a close group of
stars consisting at least of six blue stars, four of which are
possible WR~candidates (Vacca \textit{et al.} 2007). The cluster
4-1 identified by Hunter (2001) near the brightest, northern part
of the shell nebula HL~111 is the richest and youngest in the
galaxy; it also includes the close group of WR~stars M24. In the
middle of the cavity inside the shell of HL~111 is the cluster
4-2. The radius of 4-1 at its half-brightness level is 5~pc, while
the radius of 4-2 is 3~pc; the ages of both clusters are three to
four million years (Hunter 2001). (We use the term ``cluster''
following Hunter (2001); in fact, it was demonstrated by Hunter
(2001) that most stellar groups identified in IC~10 were
OB~associations, comparable in the number and concentration of
their stars to average associations in the Galaxy and Magellanic
Clouds.)

The stellar density in the field around M24 is very high; Vacca
\textit{et al.} (2007) found stars from two star-formation bursts
in this central region: one burst younger than 10 million years,
and another older burst with $t=150{-}500$~million years.
According to Vacca \textit{et al.} (2007), young blue stars in the
region of~M24 form an association, with a size at the
half-brightness level of about 6~pc. This is actually the cluster
4-1 discovered by Hunter (2001): because the coordinates of the
star M24-A in Vacca \textit{et al.} (2007) are in error by about
$2.4''$ (Vacca 2009), the relative positions of the association
and the cluster 4-1 in Fig.~13 of Vacca \textit{et al.} (2007) are
incorrect, and in fact, these objects coincide. Old red stars are
uniformly distributed in the field. The location of the two
populations of different ages along the line of sight also
differs: Vacca \textit{et al.} (2007) concluded based on the color
excesses that the blue stars forming the association were
situated at the front boundary of the high-density cloud, in
front of the population of old stars. We show below that the
results of our kinematic study do not unambiguously confirm this
location for the young stars.

According to Vacca \textit{et al.} (2007),~19 blue stars of
spectral type~O8 or earlier identified in the central cluster
around M24, including four WR~candidates, provide an ionizing
flux of about $10^{51}$~photons/s. With the electron density of
$60{-}100$~cm$^{-3}$ we measured in the region, this flux is
sufficient to explain the observed bright optical emission of the
gas in HL~111. Young stars of the cluster 4-2 provide an
additional source of ionizing radiation.

Interaction of the winds from the stars of M24 and the two
clusters 4-1 and 4-2 with the ambient gas can also explain the
observed small velocity deviations at the maximum of the H$\alpha$
and [SII] lines, as well as the weak high-velocity features in
their wings. According to our $P/V$ diagrams (Fig. 2 displays some
of them as examples), the observed shifts of the mean velocity at
the maxima of the H$\alpha$ and [SII] lines indicate a possible
systematic expansion velocity of the bright shell HL~111 not
exceeding 10--15~km/s. At the same time, we are the first to
detect weak high-velocity features in the H$\alpha$ line wings in
the cavity inside HL~111, which are displaced towards positive
velocities by up to 100--120~km/s relative to the velocity at the
line maximum. The features in the blue wings of the H$\alpha$ line
are somewhat weaker here.Weak features displaced towards positive
and negative velocities are also observed in the
[SII]~$\lambda6717$~\AA\ line (Fig.~4). The location of these
high-velocity motions in the image of HL~111 suggests they are
associated with winds from the stars of M24 belonging to the
cluster 4-1, as well as from the stars of the cluster 4-2. Note
that the most prominent features in the red wings of the line are
revealed near the cluster 4-2.

Generally speaking, the weak high-velocity features shifted
towards positive velocities, like those we observe most clearly
in the cavity inside HL~111, should be associated with the
receding side of the wind-swept, blister-type shell, at the far
boundary of the high-density cloud. However, if, as Vacca
\textit{et al.} (2007) believe, the young blue stars of M24 and
the surrounding cluster are located at the front boundary of the
high-density cloud, the receding side of the shell should be
brighter and have a lower velocity than the approaching side.
Vacca \textit{et al.} (2007) favor such a localization because of
the lower reddening of the young M24 stars and the cluster
compared to old stars in the field. However, we cannot rule out
the possibility that M24 and the parent young star cluster are
located at the back side of the HI layer, and that their low
reddening is due to strong winds from several WR~stars that have
swept away the ambient gas, considerably reducing the local
extinction.

If the stars of M24 are located closer to the far side of the
high-density gas layer, it becomes easy to explain the observed
velocity pattern. In this case, bright features of ionized-gas
lines with low shifts correspond to the bright approaching side of
a blister-type shell in a high-density medium, whereas weak
high-velocity features correspond to the blister's receding side
in a low-density medium. This model is favored by the results of
our studies of the neutral-gas kinematics. The fact that the
approaching side of the HI shell around HL~111 is brighter than
the receding side in virtually all of our 21~cm $P/V$ diagrams
indicates a lower density of the ambient neutral gas at the far
side of the envelope.

However, if the localization suggested by Vacca \textit{et al.}
(2007) is correct, we are forced to assume that the observed high
velocities in the red line wings characterize a shock induced by
the stellar wind, which propagates in a rarefied medium between
high-density gas clumps in the cloud. In fact, the distribution
of the HI column density displayed in Fig.~5 provides evidence
for a non-uniform ``clumpy'' structure of the ambient neutral
hydrogen; the same is also indicated by the $P/V$ diagrams in
Fig.~6. In this case, the weak high-velocity features in the blue
line wings could be associated with a stellar-wind-induced shock
propagating in a low-density medium at the approaching side of
the swept shell.

In both described schemes for the relative location of the HI
layer and the young stars, the kinetic energy of the wind from
the four WR stars detected here and from the two young clusters,
4-1 and 4-2, is more than sufficient to form the bright shell
nebula HL~111 35~pc in size and the expanding HI~shell around it,
even taking into account the reduced mass-loss rate in
low-metallicity galaxies. (According to Bouret (2003), the mass
loss rate depends on metallicity as $\dot M \propto Z^{m}$, where
$0.5<m< 1.0$. For $Z\simeq 0.2Z\odot$ in the galaxy IC~10
(Lozinskaya \textit{et al.} 2009, and references therein), the
stellar winds become weaker by a factor of two to five.)

We found the electron density in the nebula HL~111 using the
intensity ratio of the [SII] doublet lines from our MPFS
observations. In its brightest part, HL~111c, the density reaches
$N_{e} \simeq 60 {-} 100$~cm$^{-3}$, decreasing to $N_{e}\simeq
20 {-} 50$~cm$^{-3}$ in weaker regions. These estimates agree
with those based on slit-spectrograph observations by Lozinskaya
\textit{et al.} (2009).

In the classical theory of Castor \textit{et al.} (1975) and
Weaver \textit{et al.} (1977), the time required for a stellar
wind to sweep out the HL~111 shell, which is 35~pc in size and
expanding at a rate of $15{-}20$~km/s, is $t = (0.5 {-}
0.7)\times 10^{6}$ yrs, in agreement with the duration of the
WR~stage. Here, the required energy of the stellar wind in an
unperturbed medium with a density of $n_{0} \simeq 10$~cm$^{-3}$
reaches $L_{w}=(0.4{-}1) \times 10^{37}$~erg/s. For a metallicity
of $Z\simeq 0.2Z\odot$, this energy corresponds to the wind from
the four WR~stars belonging to M24. An additional source of
mechanical energy is the wind from the stars of the two clusters
4-1 and 4-2. In both spatial schemes, the bright line-profile
features at low velocities are due to gas behind the front of the
shock propagating in a high-density medium, while the weak
features in the wings at high velocities of about 100--120~km/s
are due to the emission of gas behind the front of the shock in a
low-density medium.

We can compare the structure and kinematics of the ionized and
neutral gas in the neighborhood of the center of recent star
formation M24 and HL~111 to an isolated star-forming region
studied in detail in the wing of the Small Magellanic Cloud: the
young cluster NGC~602 and the related nebula N90. Observations in
the 21~cm line and the optical provide evidence that the cluster
NGC~602 was formed at the periphery of an HI cloud, and that the
nebula N90 is a blister at its edge (Nigra \textit{et al.} 2008).
Here, in contrast to the M24 -- HL~111 region in IC~10, the
blister and cloud edge are situated in the plane of the sky,
considerably simplifying interpretation of the observations. The
age of NGC~602 is about four to five million years (Hutchings
\textit{et al.} 1991; Carlson \textit{et al.} 2007), the same as
the age of the clusters 4-1 and 4-2. The two elongated groups of
young stars detected here probably indicate propagating star
formation (Gouliermis \textit{et al.} 2007). Parameters of the
HI~cloud are also very similar to those of the star-forming
region in IC~10: its column density is
$N({\textrm{HI}})=(3-4)\times10^{21}$~cm$^{-2}$, its diameter is
220~pc, and its mass is about $2.7\times10^{5}~M_{\odot}$. The
ionized shells N90 and HL~111 in IC~10 are also similar: the size
of N90 in H$\alpha$ is 40--50 pc, with ionized-gas velocities
between 175 and 184 km/s. Thus, the nebula N90, like the bright
shell HL~111, is kinematically very ``quiet'' judging from the
velocities at the line maximum. Unlike HL~111, the region of N90
reveals no high-velocity features in the line profile, which can
be explained by the spatial orientation of the ``blister''.

{\textbf{The nebula HL~106}} is located towards the molecular
cloud that is brightest in the CO line. Figure~1b shows the
structure of this cloud, overlaid on the H$\alpha$ image of the
region. (We plotted a combined map of the cloud by putting
together maps for its individual components: B11a, B11b, B11c, and
B11d from Fig.~7 in Leroy \textit{et al.} 2006.)

The distribution of interstellar extinction in the field can
clarify the relative position of the high-density cloud and the
nebula HL~106 in the line of sight. The slit spectrogram PA~331
(the positions ${-}19\div {-}10$ in Fig.~1 of Lozinskaya
\textit{et al.} 2009) reveals a color excess at the eastern edge
of HL~106 and the CO~cloud B11b from a comparison of the observed
line-flux ratio, H$\alpha:\textrm{H}\beta:\textrm{H}\gamma$, to
the theoretical ratio H$\alpha:\textrm{H}\beta:\textrm{H}\gamma
=2.86:1.00:0.47$ (Aller 1984). Only here does the color excess
reach its highest value, $E(B-V) = 1.3^m\pm 0.15 ^{m}$, however
the values observed in the center of the region crossed with the
slit are average for the region, $E(B-V)=0.8^m {-}0.9^{m}$.

In the present study, we used the same lines from our MPFS
observations to estimate the color excess in the western part of
HL~106 and the cloud B11b (Fig.~1b). The color excess at the
bright arc near the WR~star R10 is $E(B-V) = 0.9^m{-}1.0 ^{m}$,
then decreases towards the shell center to values of about
$E(B-V)=0.7^{m}$, and further increases to $E(B-V)=1.1^{m}$ near
the compact HII region HL~106a in the central part of the CO
cloud.

The extinction for the six brightest compact HII regions in the
area was determined from observations in the Br$\gamma$ line by
Borissova \textit{et al.} (2000). For the compact region Br5,
coincident with HL~106a, $E(B-V) = 2.18^{m}$; the color excesses
in three other Br$\gamma$ sources of the star-forming complex are
within $1.45^m{-} 1.88^{m}$. The $E(B-V)$ values derived in this
study differ somewhat from the data of other authors, however it
indicate a higher extinction in the central part of the CO~cloud
B11b. (Note the systematic shift of the coordinates of the
Br$\gamma$ sources in Table~2 of Borissova \textit{et al.} (2000);
for this reason, their positions in Fig.~16 of Hunter (2001) are
incorrect. We independently determined the positions of these
compact sources from the Br$\gamma$ isophotes by identifying them
in the H$\alpha$ image of the galaxy.)

Thus, the existing estimates indicate a slight increase in the
extinction in the region of the CO cloud B11b, by about
$0.1^m{-}0.3^{m}$, corresponding, for solar abundances, to an
increase in the column density by $\Delta N({\textrm
HI})\simeq(0.5{-}1.5)\times 10^{21}$~cm$^{-2}$. Even taking into
account the low metallicity of IC~10, this is a small fraction of
the column density towards the high-density cloud, which reaches
$N\textrm{(HI)}\simeq 10^{22}$~cm$^{-2}$ (Wilcots and Miller
1998). The total density of neutral and molecular hydrogen derived
from observations of CO emission in the region is
$N({\textrm{H}})\simeq 2.8 \times10^{22}$~cm$^{-2}$  (Leroy
\textit{et al.} 2006; Bolatto \textit{et al.} 2000). Thus, it may
be that the optical nebula HL~106 is not behind the dense layer of
clouds, but is instead partially embedded in it. Such a relative
position of HL~106 and the molecular cloud is also favored by the
fact that the far side of the HI shell we detected around HL~106
is brighter than the front side (Figs.~6a,~6b).

Note that the cloud B11b is probably physically associated with
the optical nebula. First of all, the radial velocity of the
CO~cloud B11b ($V= {-}331$~km/s according to Leroy \textit{et al.}
2006) exactly coincides with the velocity of the ionized gas in
the shell  HL~106 (Figs.~3,~4). The most important thing, however,
is that the brightest southern arc of HL~106 exactly follows the
edge of the B11b cloud (Fig.~1b). This perfect coincidence cannot
be due to chance, and testifies to a physical association between
the thin ionized shell and the molecular cloud B11b. It is
possible that the shell HL~106 exactly outlining the CO cloud B11b
was formed by photo-dissociation of molecular gas at the edge of
this cloud and ionization by UV~radiation from the WR stars R2 and
R10 and the two clusters 4-3 and 4-4 (Fig.~1). The age of these
clusters estimated in Hunter (2001) is about 20-–30~million years.

Our study of the kinematics of ionized and neutral gas in the
nebula HL~106 has demonstrated that we observe here a pattern
similar to that for HL~111. The velocities derived from the maxima
of the H$\alpha$ and [SII] lines and characterizing the brightest
clumps in the nebula show slight variations and give a mean
expansion rate of the shell of 10-–15~km/s. Weak features are seen
in the line wings in HL~106, at levels of about $2\%{-}10\%$ of
the peak intensity, at velocities ${-}260 \div {-}250$~km/s and
${-}450 \div {-}420$~km/s.

The outer neutral-gas shell around HL~106 also has a size and
expansion rate similar to those for the HI shell around HL~111.

As in HL~111, it is easy to establish that the winds from the WR
stars R2 and R10 and from the clusters 4-3 and 4-4 supply a
mechanical energy sufficient to form the shell.

{\textbf{The kinematics of gas in the star-formation complex}}
results from the complicated structure of the ambient neutral gas
in the region, the presence of molecular clouds and numerous
sources of stellar wind. In this situation, it is difficult to
expect regular morphology and kinematics of the ionized gas in the
region of violent star formation.

As was shown using the example of the two shells HL~111 and
HL~106, the interaction of the WR stars and massive blue stars of
the clusters situated here with the ambient HI cloud, possibly
influenced by the previous stellar generation, can explain the
observed slight deviations of the mean velocity at the maxima of
the H$\alpha$ and [SII] lines, as well as the weak high-velocity
features in their wings.

The column density of neutral hydrogen is lower in the western
part of the region between HL~111 and HL~106 (Fig.~5). As it was
demonstrated in Section~3.1, high-velocity features are also
observed here in the H$\alpha$ and [SII] $\lambda6717$~\AA\ line
profiles. The effects of the winds from the three WR~stars located
here M13, M12, and M14 can explain the presence of high-velocity
wings in this region. The winds from these stars produce a clear
pattern of interaction with the ambient medium: weak arc and ring
structures are observed at this position (Fig.~1). The two
brighter, compact HII regions, HL~111a and HL~111b, are probably
the highest-density clumps of the HI cloud ionized by these
stars. Here, at the western edge of the cloud, the winds from WR
stars disrupt the ``parent'' HI cloud most strongly because of
the reduced neutral-gas density.

As was demonstrated in Section 3.1, weak high-velocity features in
the wings of both lines are also observed in other regions of the
star-formation complex. It is easy to understand the abundance of
high-velocity motions in the complex, given the high density of
stellar-wind sources and the non-uniform ambient interstellar
medium.

The study of the large-scale kinematics of neutral hydrogen in
IC~10 presented by Wilcots and Miller (1998) revealed extended
shells of neutral gas about $1'{-}1.5'$ in size (cf. Fig.~9 in
Wilcots and Miller 1998). Our re-analysis of the 21-cm VLA
observations of IC~10 was aimed at a more detailed study of the
``local'' structure and kinematics of HI in the star-formation
complex. As a result, we were able to identify two local neutral
shells about $20''{-}30''$ in size around the two brightest
optical nebulae, HL~111 and HL~106, to measure their expansion
rates, and to suggest possible sources of the kinetic energy
needed to form them.

We also detected neutral-gas motions with velocities up to
30~km/s on scales of $20''- 40''$ to the east and west of the
complex of bright nebulae. WR~stars and young clusters that can
initiate such motions are located here.

\section{CONCLUSIONS}

We have used our observations of the galaxy IC~10 using the SAO
6-m telescope with the SCORPIO focal reducer in the Fabry-–Perot
interferometer mode and with the MPFS panoramic spectrograph to
study the structure and kinematics of ionized gas in the central
region of intense recent star formation. We have also used
archival 21-cm VLA observations to examine the structure and
kinematics of neutral gas in the area. The velocity of the ionized
gas in the region of violent star formation, which contains the
brightest nebulae, HL~111 and HL~I06, varies between ${-}315$ and
${-}350$~km/s, with a mean velocity of about ${-}330$~km/s. The
neutral gas in the region has the same velocity. These findings
agree with the H$\alpha$ measurements of Thurow and Wilcots~(2005)
(whose observations cover one-third of the bright star-forming
region we have studied) and with the estimates of Wilcots and
Miller (1998) based on 21~cm data.

We have demonstrated that the mean expansion rate of the bright
shells HL~111 and HL~106 do not exceed 10-–15 km/s. We have also
detected for the first time high-velocity wings of the H$\alpha$
and [SII] emission lines at velocities between ${-}280$ and
${-}240$~km/s and between ${-}450$ and ${-}410$~km/s in the inner
cavity of the nebula HL~111, as well as in other parts of the
violent star-formation complex.

The archival 21-cm VLA observations revealed two neutral shells
about $20''{-}30''$ in size around the nebulae HL~111 and HL~106.
The mean expansion rate of both HI shells is 15–-20 km/s, and the
highest expansion velocity of neutral-gas clumps reaches 25–-30
km/s. We have suggested possible sources of the kinetic energy
needed to form the HII shells and the surrounding HI shells.

We also detected neutral-gas motions with velocities up to
30~km/s on scales of $20''{-}40''$ to the east and west of the
complex of bright nebulae. The WR~stars and young clusters located
here are able to initiate such motions.

We have discussed a spatial scheme for the interaction between
the stellar and gaseous populations in the central violent
star-formation complex in the galaxy.

\section*{ACKNOWLEDGEMENTS}

This study was supported by the Russian Foundation for Basic
Research (project code 07-02-00227). The study was based on
observations with the 6-m telescope of the Special Astrophysical
Observatory, which is funded by the Ministry of Education and
Science (registration number 01-43). The authors thank W.D.~Vacca
for discussions of the fit of coordinates of the star M24-A in his
paper; E.~Wilcots for providing the data cube of the HI
observations; and N.Yu.~Podorvanyuk for providing one of the codes
used to reduce these data. We have made use of the NASA/IPAC
Extragalactic Database (NED), which is operated by the Jet
Propulsion Laboratory, California Institute of Technology, under
contract with the National Aeronautics and Space Administration
(USA).

\bigskip
\centerline{REFERENCES}
\bigskip

\noindent Afanasiev,~V.~L., Dodonov,~S.~N., and Moiseev,~A.~V.,
in: {\emph{Stellar dynamics: from classic to modern}}, eds
Osipkov,~L.~P., Nikiforov,~I.~I. (Saint-Petersburg:
Saint-Petersburg Univ. Press, 2001), p.~103.

\noindent Aller,~L.~H., {\emph{Physics of thermal gaseous
nebulae}} (D.~Reidel Publ. Comp., 1984).

\noindent Begum,~A., Chengalur,~J.~N., Karachentsev,~I.~D.,
{{\textit{et al.}}}, Monthly Not. Roy. Astron. Soc. \textbf{365},
1220 (2006).

\noindent Bolatto,~A.~D., Jackson,~J.~M., Wilson,~C.~D., and
Moriarty-Schieven,~G., Astrophys. J. \textbf{532}, 909 (2000).

\noindent Borissova,~J., Georgiev,~L., Rosado,~M., {{\textit{et
al.}}}, Astron. and Astrophys. \textbf{363}, 130 (2000).

\noindent Bouret,~J.~C., Lanz,~T., Hillier,~D.~J., {{\textit{et
al.}}}, Astrophys. J. \textbf{595}, 1182 (2003).

\noindent Carlson,~L.~R., Sabbi,~E., Sirianni,~E., {{\textit{et
al.}}}, Astrophys. J. (Letters) \textbf{665}, L109 (2007).

\noindent Castor,~J., McCray,~R., and Weaver,~H., Astrophys. J.
(Letters) \textbf{200}, L107 (1975).

\noindent Chyzy,~K.~T., Knapik,~J., Bomans,~D.~J., {{\textit{et
al.}}}, Astron. and Astrophys. \textbf{405}, 513 (2003).

\noindent Crowther,~P.~A., Drissen,~L., Abbott,~J.~B.,
{{\textit{et al.}}}, Astron. and Astrophys. \textbf{404}, 483
(2003).

\noindent Gil de Paz,~A., Madore,~B.~F., and Pevunova,~O.,
Astrophys. J.~Suppl. Ser. \textbf{147}, 29 (2003).

\noindent Gouliermis,~D.~A., Quanz,~S.~P., and Henning,~T.,
Astrophys. J. \textbf{665}, 306 (2007).

\noindent Hodge,~P. and Lee,~M.~G., Publs Astron. Soc. Pacif.
\textbf{102}, 26 (1990).

\noindent Hunter,~D.~A., Astrophys. J. \textbf{559}, 225 (2001).

\noindent Hutchings,~J.~B., Cartledge,~S., Pazder,~J., and
Thompson,~I.~B., Astron. J. \textbf{101}, 933 (1991).

\noindent Kim,~M., Kim,~E., Hwang,~N., {{\textit{et al.}}},
Astrophys. J. \textbf{703}, 816 (2009); e-Print
arXiv:astro-ph/0907.4844.

\noindent Kniazev,~A.~Y., Pustilnik,~S.~A., and Zucker,~D.~B.,
Monthly Not. Roy. Astron. Soc. \textbf{384}, 1045 (2008).

\noindent Leroy,~A., Bolatto,~A., Walter,~F., and Blitz,~L.,
Astrophys. J. \textbf{643}, 825 (2006).

\noindent Lozinskaya,~T.~A. and Moiseev,~A.~V., Monthly Not. Roy.
Astron. Soc. \textbf{381}, 26L (2007).

\noindent Lozinskaya,~T.~A., Moiseev,~A.~V., Podorvanyuk,~N.~Yu.,
and Burenkov,A.~N., Pis'ma Astron. Zh. \textbf{34}, 243 (2008).

\noindent Lozinskaya,~T.~A., Egorov,~O.~V., Moiseev,~A.~V., and
Bizyaev,~D.~V., Pis'ma Astron. Zh. \textbf{35}, 811 (2009).

\noindent Massey,~P., Armandroff,~T.~E., and Conti,~P.~S.,
Astron. J. \textbf{103}, 1159 (1992).

\noindent Massey,~P. and Holmes,~S., Astrophys. J. (Letters)
\textbf{580}, L35 (2002).

\noindent Massey,~P., Olsen,~K., Hodge,~P., {{\textit{et al.}}},
Astron. J. \textbf{133}, 2393 (2007).

\noindent Moiseev,~A.~V., Bull. Spec. Astophys. Observ.
\textbf{54}, 74 (2002); e-Print arXiv:astro-ph/0211104.

\noindent Moiseev,~A.~V., Egorov,~O.~V., Astrofiz. Byull.
\textbf{63}, 193 (2008).

\noindent Nigra,~L., Gallagher,~J.~S., Smith,~L.~J., {{\textit{et
al.}}}, Publs Astron. Soc. Pacif. \textbf{120}, 972 (2008);
e-Print arXiv:astro-ph/0808.1033.

\noindent Richer,~M.~G., Bullejos,~A., Borissova,~J., \textit{et
al.}, Astron. and Astrophys. \textbf{370}, 34 (2001).

\noindent Royer,~P., Smartt,~S.~J., Manfroid,~J., and Vreux,~J.,
Astron. and Astrophys. \textbf{366}, L1 (2001).

\noindent Sanna,~N., Bono,~G., Stetson,~P.~B., {{\textit{et
al.}}}, Astrophys. J. (Letters) \textbf{688}, L69 (2008a).

\noindent Sanna,~N., Bono,~G., Monelli,~M., {{\textit{et al.}}},
Mem. Soc. Astron. Ital. \textbf{79}, 747 (2008b).

\noindent Sharina,~M.~E., Chandar,~R., Puzia,~T.~H.,
Coudfrooij,~P., and Davoust,~E., Monthly Not. Roy. Astron. Soc.
in press (2010); e-Print arXiv:astro-ph/1002.2144v1

\noindent Thurow,~J.~C. and Wilcots,~E.~M., Astron. J.
\textbf{129}, 745 (2005).

\noindent Tikhonov,~N.~A. and Galazutdinova,~O.~A., Pis'ma Astron.
Zh. \textbf{35}, 829 (2009).

\noindent Vacca,~W.~D., private commun. (2009).

\noindent Vacca,~W.~D., Sheehy,~C.~D., and Graham,~J.~R.,
Astrophys. J. \textbf{662}, 272 (2007).

\noindent Weaver,~H., McCray,~R., Shapiro,~P., and Moore,~R.,
Astrophys. J. \textbf{218}, 377 (1977).

\noindent Wilcots,~E.~M. and Miller,~B.~W., Astron. J.
\textbf{116}, 2363 (1998).

\noindent Yang,~H. and Skillman,~E.~D., Astron. J. \textbf{106},
1448 (1993).

\noindent Zucker,~D.~B., Bull. Amer. Astron. Soc. \textbf{32},
1456 (2000).

\noindent Zucker,D.~B., Bull. Amer. Astron. Soc. \textbf{34},
1147 (2002).

\end{document}